\documentstyle[11pt,epsfig]{article}
\title{\bf Thermodynamics of noncommutative de Sitter spacetime}
\author{ B. Vakili\thanks{email: b-vakili@sbu.ac.ir}, N. Khosravi
\thanks{email: n-khosravi@sbu.ac.ir} and H. R. Sepangi
\thanks{email: hr-sepangi@sbu.ac.ir}
\\ {\small Department of Physics, Shahid Beheshti University, Evin, Tehran 19839, Iran}}
\begin{document}
\maketitle 
\begin{abstract}
We study the effects of noncommutativity of spacetime geometry on
the thermodynamical properties of the de Sitter horizon. We show
that noncommutativity results in modifications in temperature,
entropy and vacuum energy and that these modifications are of
order of the Planck scale, suggesting that the size of the
noncommutative parameter should be close to that of the Planck. In
an alternative way to deal with noncommutativity, we obtain a
quantization rule for the entropy. Since noncommutativity in
spacetime geometry modifies the Heisenberg algebra and introduces
the general uncertainty principle, we also investigate the above
problem in this framework.
\vspace{5mm}\newline Keywords: de Sitter spacetime;
Noncommutativity; Generalized Uncertainty
Principle.\vspace{5mm}\newline PACS numbers: 04.60.-m, 04.60.Ds,
04.60.Kz\vspace{.5cm}
\end{abstract}
\section{Introduction}
The black hole thermodynamics formulation, first introduced by
Bekenstein \cite{1} and a few years later by Hawking \cite{2}, is
based on applying quantum field theory to the curved spacetime of
a black hole. According to this formalism a black hole emits
radiation with a temperature
\begin{equation}\label{A}
T_{BH}=\frac{\hbar \kappa}{2\pi k_B c},
\end{equation}
where $\kappa$ is the surface gravity at the black hole horizon
and $k_B$ is the Boltzmann constant. In the case of a
Schwarzschild black hole the temperature can be written as
\begin{equation}\label{B}
T_{BH}=\frac{\hbar c^3}{8\pi Gk_Bm},
\end{equation}
with an entropy, given by
\begin{equation}\label{C}
S_{BH}=\frac{1}{4}\frac{k_Bc^3}{G
\hbar}A,
\end{equation}
where $A$ is the area of the black hole event horizon. The Hawking
radiation of a black hole is due to the random processes in the
quantum fields near the horizon. The mechanism of this thermal
radiation can be explained in terms of pair creation in the
gravitational potential well of the black hole \cite{3}. The
conclusions of the above works is that the temperature of a black
hole is proportional to the surface gravity and that the area of
its event horizon plays the role of its entropy. In this scenario
the black hole is akin to a thermodynamical system obeying the
usual thermodynamics laws, often called the laws of black hole
mechanics, first formulated by Hawking \cite{2}.

In more recent times, this issue has been at the center of
concerted efforts in the hope of describing and making clear
various aspects of the problem  still remaining unclear, for a
review see \cite{4}. With the birth of string theory \cite{5} as a
candidate for quantum gravity and loop quantum gravity \cite{6}, a
new window was opened to the problem of black hole radiation. This
was because the nature of black hole radiation is such that
quantum gravity effects can not be neglected \cite{7}. Motivated
by sting theory, noncommutative spaces \cite{8} and the
Generalized Uncertainty Principle (GUP) \cite{9} have been studied
extensively and applied to black hole physics, leading to some
modifications of the Hawking theory which were comparable with
those that came from string theory and loop quantum gravity
\cite{10}-\cite{16}.

An interesting feature of the black hole thermodynamics theory is
the close connections between the black hole event horizon and
cosmological horizon in so far as the thermodynamical properties are
concerned. This issue was first introduced by Gibbons and Hawking in
\cite{17} where they extended the black hole thermodynamics to the
cosmological models with a positive (repulsive) cosmological
constant. Indeed, they showed that in analogy with a black hole, the
area of the event horizon in these models can play the role of
entropy or lack of information on the regions inaccessible to an
observer. Also, one can associate a surface gravity, $\kappa$, to
the event horizon which can be interpreted as temperature. Applying
the same procedure to the cosmological event horizon, they showed
that laws very similar to zeroth, first and second laws of black
hole mechanics are obeyed. Particulary, the particle creation with a
thermal spectrum would also occur in these cosmological models, as
is shown in \cite{17} and \cite {Ji} for a de Sitter model.
Recently, it has been shown that the Friedmann equation describing
the dynamics of an accelerated expanding universe can be obtained by
applying the first law of thermodynamics to the apparent horizon of
the FRW universe. This issue is not only investigated in the usual
Einstein-Hilbert framework but in the case of scalar-tensor type
$f(R)$ gravity theories like that of the Gauss-Bonnet and the more
general Lovelock \cite{18}-\cite{21}.

In this paper, motivated by the works done in the context of
noncommutative black hole physics, we study the effects of
noncommutativity on the thermodynamical properties of de Sitter
spacetime as the simplest model with a cosmological horizon. As is
well known from the work of Gibbons and Hawking \cite{17}, this
spacetime has an apparent horizon which behaves like the black
hole horizon as far as thermal radiation is concerned. Since this
phenomena is described quantum mechanically, the quantum
gravitational effects can not be neglected and this paves the way
for our physical motivation to deal with this problem. Since there
is not a unique approach in dealing with noncommutativity in such
models, we consider different approaches and compare the results.

The organization of the paper is as follows: in section 2, we review
the thermodynamical properties of the de Sitter space. In section 3,
to introduce noncommutativity, we directly replace the coordinates
with their noncommutative counterparts and obtain new
thermodynamical variables which are different from their commutative
version. In section 4, we deal with noncommutative spaces in an
alternative way by considering the horizon equation as a constraint
and obtain a quantization rule for entropy. Section 5 deals with the
same problem in the framework of GUP. Finally, we summarize the
conclusions in section 6. In what follows, we work in units where
$c=\hbar=G=k_B=1$.
\section{Thermodynamics of the de Sitter spacetime: commutative case}
The simplest solution to Einstein equations with a cosmological
constant, $R_{\mu \nu}-\frac{1}{2}Rg_{\mu\nu}+\Lambda
g_{\mu\nu}=8\pi T_{\mu\nu}$, is the de Sitter spacetime which is a
solution of the above field equations with $T_{\mu\nu}=0$
\cite{22}. The corresponding metric in a static form can be
written as
\begin{equation}\label{D}
ds^2=-\left(1-\frac{\Lambda}{3}r^2\right)dt^2+
\left(1-\frac{\Lambda}{3}r^2\right)^{-1}dr^2+r^2\left(d\vartheta^2+
\sin^2 \vartheta d\phi^2\right).
\end{equation}
This metric has an apparent singularity at
$r_h=\sqrt{3/\Lambda}$.\footnote{We consider the positive
cosmological constant only. It is clear that for a negative
cosmological constant, the metric does not have any singularity.
In general, models with negative cosmological constant do not have
an event horizon \cite{17}.} This singularity is like the
coordinate singularity associated with horizon in the
Schwarzschild spacetime, and as is well known, there are other
coordinate systems for which this type of singularity is removed
\cite{23}. Thus, such a singularity is of the horizon type,
meaning that the light cone tips over for $r>\sqrt{3/\Lambda}$ and
$\partial/\partial r$ becomes timelike while $\partial/\partial t$
becomes spacelike. This means that one can send signals outside
the horizon through the $r_h=\sqrt{3/\Lambda}$ surface but can
never receive information from there, {\it i. e.} the de Sitter
horizon is a one-way membrane, from inside to the outside.

As we mentioned in the introduction, the similarity between the
black hole horizon and the de Sitter horizon is the main
motivation to assign a temperature and an entropy to the horizon
in the de Sitter space. In spite of their different nature, one
being a black hole horizon and the other a cosmological one as is
shown in \cite{17}, they have many of the same thermodynamical
properties.

In this section, we briefly review the thermodynamics of the de
Sitter horizon in a commutative spacetime with the usual canonical
commutation relations prevailing, that is
\begin{equation}\label{E}
[x_i,x_j]=0,\hspace{.5cm}
[x_i,p_j]=i\delta_{ij},\hspace{.5cm}[p_i,p_j]=0.
\end{equation}
In the classical limit the quantum mechanical commutators should be
replaced by the classical Poisson brackets as $[P,Q]\rightarrow i
\{P,Q \}$. As we indicated above, the de Sitter spacetime (\ref{D})
has a horizon at $r_h=\sqrt{3/\Lambda}$ with the following area
\begin{equation}\label{F}
A=r_h^2\int_0^{2\pi}d\phi \int_0^{\pi}\sin^2\vartheta
d\vartheta=4\pi r_h^2=\frac{12 \pi}{\Lambda}.
\end{equation}
This horizon is a null surface, that is, vectors normal to this
surface are null. Thus, if we consider the Killing vector ${\bf
\xi}$, generating time translations (${\bf \xi}=\partial/\partial
t$), it is normal to the horizon ($\xi_{\mu}\xi^{\mu}=0$).
Therefore, $\xi_{\mu}\xi^{\mu}$ is constant on the horizon and
thus $\nabla^{\alpha}(\xi_{\mu}\xi^{\mu})$ is also normal to the
horizon. If one normalizes ${\bf \xi}$ to have unit magnitude at
the origin \cite{17}, one can define a surface gravity for the
horizon, such that on the horizon
\begin{equation}\label{G}
\nabla^{\alpha}(\xi_{\mu}\xi^{\mu})=2\kappa
\xi^{\alpha}.
\end{equation}
For a diagonal metric in spherical coordinates whose components
are dependent on $r$ only, it is easy to show that
\begin{equation}\label{AA}
\kappa=\sqrt{-\frac{1}{4}g^{rr}g^{tt}\left(\frac{\partial
g_{tt}}{\partial r}\right)^2},
\end{equation}
evaluated at the horizon. For the de Sitter space (\ref{D}), this
gives
\begin{equation}
\label{H}\kappa=\sqrt{\frac{\Lambda}{3}}.
\end{equation}
According to the Gibbons and Hawking analysis of particle creation
in de Sitter spacetime \cite{17}, one can assign temperature and
entropy to this spacetime, using equations (\ref{A})-(\ref{C}),
(\ref{F}) and (\ref{H}), leading to
\begin{equation}\label{I}
T_{GH}=\frac{1}{\beta_{GH}}=\frac{1}{2\pi}\sqrt{\frac{\Lambda}{3}},
\end{equation}
\begin{equation}\label{J}
S_{GH}=\frac{3\pi}{\Lambda}.
\end{equation}
This temperature is associated with an isotropic background of
thermal radiation measured by a freely moving observer in de
Sitter space. One can describe this thermal radiation in terms of
pair creation in the gravitational potential well of the de Sitter
spacetime. Outside the horizon and near it, there are particle
states which are energetically allowed to create a pair of
particles. One such particle goes to infinity and the other can
tunnel through the horizon and become absorbed by an observer in
the de Sitter space. The rate of such particle emission is
calculated in \cite{17} where it is shown that the result is
similar to radiation from a black body with temperature
$T=\frac{1}{2\pi}\sqrt{\Lambda/3}$.

Applying the first law of thermodynamics
\begin{equation}\label{L}
dE_{GH}=T_{GH}dS_{GH},
\end{equation}
to the horizon one can evaluate the vacuum energy of the de Sitter
space by integration. Using equations (\ref{I}) and (\ref{J}), we
obtain
\begin{equation}\label{K}
E_{GH}=\sqrt{\frac{3}{\Lambda}}=\frac{1}{2\pi}\beta_{GH}.
\end{equation}
There is also an alternative way to derive the above results using
statistical methods. In this method, we first express the de
Sitter metric in Kruskal coordinates \cite{17}
\begin{equation}\label{M}
ds^2=\frac{3}{\Lambda}(UV-1)^{-2}\left[-4dUdV+(UV+1)^2(d\vartheta^2+\sin^2\vartheta
d\phi^2)\right],
\end{equation}
where
\begin{equation}\label{N}
r=\sqrt{\frac{3}{\Lambda}}(UV+1)(1-UV),\hspace{.5cm}\exp\left(2\sqrt{\frac{\Lambda}{3}}t\right)=-VU^{-1}.
\end{equation}
Now, it is clear from this coordinate system that the
corresponding spacetime has an imaginary period $\beta$
\begin{equation}\label{O}
\beta=2\pi\sqrt{\frac{3}{\Lambda}},
\end{equation}
which is in agreement with relation (\ref{I}) that gives the
corresponding temperature. Now, following \cite{24}, we introduce
the partition function of the (gravitational) system through the
following path integral
\begin{equation}\label{P}
Z=\sum_n e^{-\beta E_n}=\int D[g]e^{-\hat{I}},\hspace{.5cm}\ln
Z=-\hat{I},
\end{equation}
where the summation is over the different quantum states and
$\hat{I}$ is the Euclidean action. For the de Sitter spacetime the
Euclidean action can be obtained as, see \cite{24} and \cite{25}
\begin{equation}\label{R}\hat{I}=\frac{3\pi}{ \Lambda}=\frac{1}{4
\pi}\beta^2_{GH}.
\end{equation}
Thus, we have
\begin{equation}\label{S}
\ln Z=-\frac{1}{4
\pi}\beta^2.
\end{equation}
Now, the energy and entropy of the gravitational system can be
evaluated from the usual relations
\begin{equation}\label{T}
\bar{E}=-\frac{\partial \ln Z}{\partial
\beta},\hspace{.5cm}S=\beta \bar{E}+\ln Z,
\end{equation}
with the result
\begin{equation}\label{U}
\bar{E}=\frac{1}{2\pi}\beta=\sqrt{\frac{3}{\Lambda}},\hspace{.5cm}S=\frac{1}{4\pi}\beta^2=\frac{3\pi}{\Lambda}.
\end{equation}
\section{Thermodynamics of de Sitter spacetime: noncommutative case}
Let us now concentrate on the noncommutativity concepts in the
model described above, hoping to reach modifications which are
closely related to those coming from quantum gravity or string
theory.

Noncommutativity is described by a deformed product, also known as
the Moyal product law between two arbitrary functions of position
and momenta as
\begin{equation}\label{V}
(f*_{\alpha}g)(x)=\exp\left[\frac{i}{2}\alpha^{ab}\partial^{(1)}_a
\partial^{(2)}_b\right]f(x_1)g(x_2)|_{x_1=x_2=x},\end{equation}such
that
\begin{equation}\label{X}
\alpha_{ab}=\left(%
\begin{array}{cc}
  \Theta_{ij} & \delta_{ij} \\
  -\delta_{ij} & 0 \\
\end{array}%
\right),
\end{equation}
where the $N\times N$ matrix $\Theta$ is
assumed to be antisymmetric with $2N$ being the dimension of the
phase space and represents the noncommutativity in coordinates.
With this product law, the deformed commutation relations can be
written as
\begin{equation}\label{Y}
\left[f,g\right]_{\alpha}=f*_{\alpha}g-g*_{\alpha}f.
\end{equation}
A simple calculation shows that
\begin{equation}\label{W}
\left[x_i,x_j\right]_{\alpha}=i
\Theta_{ij},\hspace{.5cm}\left[x_i,p_j\right]_{\alpha}=i
\delta_{ij},\hspace{.5cm}\left[p_i,p_j\right]_{\alpha}=0.
\end{equation}
Now, consider the following transformations
\begin{equation}\label{Z}\hat{x}_i=x_i-\frac{1}{2}\Theta_{ij}p_j,\hspace{.5cm}\hat{p}_i=p_i.
\end{equation}
It can easily be checked that if $(x_i,p_j)$ obey the usual
(Poisson) commutation relations (\ref{E}), then
\begin{equation}\label{AB}
\left\{\hat{x}_i,\hat{x}_j\right\}=\Theta_{ij},\hspace{.5cm}\left\{\hat{x}_i,\hat{p}_j\right\}=\delta_{ij},\hspace{.5cm}
\left\{\hat{p}_i,\hat{p}_j\right\}=0.
\end{equation}
These Poisson brackets are the same as the Poisson bracket
counterparts of relations (\ref{W}). Consequently, for introducing
noncommutativity, it is more convenient to work with commutation
relations (\ref{AB}) than $\alpha-star$ deformed commutation
relations (\ref{W}). It is important to note that the relations
represented by equations (\ref{W}) are defined in the spirit of the
Moyal product given above. However, in the relations defined by
equations (\ref{AB}), the variables $(\hat{x}_i,\hat{p}_j)$ obey the
usual commutation relations, so that the two sets of deformed and
ordinary commutation relations represented by (\ref{W}) and
(\ref{AB}) should be considered as distinct.

Now, consider the noncommutative de Sitter spacetime whose metric
is given by
\begin{equation}\label{AC}
ds^2=-\left(1-\frac{\Lambda}{3}\hat{r}\hat{r}\right)dt^2+
\left(1-\frac{\Lambda}{3}\hat{r}\hat{r}\right)^{-1}d\hat{r}d\hat{r}
+\hat{r}\hat{r}\left(d\vartheta^2+\sin^2 \vartheta
d\phi^2\right),
\end{equation}
where $\hat{r}$ satisfies (\ref{AB}). The horizon radius of the
noncommutative de Sitter space can be obtained from the following
condition
\begin{equation}\label{AD}1-\frac{\Lambda}{3}\hat{r}\hat{r}=0.
\end{equation}
Applying the transformations (\ref{Z}) to this condition, the
horizon of the noncommutative de Sitter space should satisfy the
following relation
\begin{equation}\label{AE}
1-\frac{\Lambda}{3}\left(x_i-\frac{1}{2}\Theta_{ij}p_j\right)
\left(x_i-\frac{1}{2}\Theta_{ik}p_k\right)=0,
\end{equation}
which leads to
\begin{equation}\label{AF}
1-\frac{\Lambda}{3}\left[r^2-\vec{\Theta}.\vec{L}+\frac{1}{4}\left(\Theta^2
p^2-(\vec{\Theta}.\vec{p})^2\right)\right]=0,
\end{equation}
where $\vec{L}=\vec{r}\times \vec{p}$ and
$\Theta_{ij}=\epsilon_{ijk}\Theta_k$. If, as is done in \cite{11},
we take $\vec{\Theta}=(0,0,\Theta)$, (which can be done by a
rotation or redefinition of coordinates), we get
\begin{equation}\label{AG}
1-\frac{\Lambda}{3}\left[r^2-\Theta
L_z+\frac{1}{4}\Theta^2(p_x^2+p_y^2)\right]=0.
\end{equation}
Since the de Sitter spacetime is non-rotating, $L_z=0$ and by
definition $\theta^2=\frac{1}{4}(p_x^2+p_y^2)\Theta^2$, we are led
to the radius of noncommutative de Sitter spacetime
\begin{equation}\label{AH}
r_{nc}=\sqrt{\frac{3}{\Lambda}-\theta^2},
\end{equation}
or taking terms to second order of the noncommutative parameter
$\theta$, we have
\begin{equation}\label{AI}
r_{nc}=r_h\left(1-\frac{1}{2}\frac{\theta^2}{r_h^2}\right).
\end{equation}
We see that the radius of the horizon in noncommutative case is
less than its commutative counterpart. The horizon area of the
noncommutative de Sitter space defined as $A_{nc}=4\pi r_{nc}^2$
yields
\begin{equation}\label{AJ}
A_{nc}=A\left(1-\frac{4\pi
\theta^2}{A}\right).
\end{equation}
Thus, the noncommutative entropy corresponding to the
noncommutative space reads $S_{nc}=\frac{1}{4}A_{nc}$, with the
result
\begin{equation}\label{AL}
S_{nc}=S_{GH}\left(1-\frac{\pi \theta^2}{S_{GH}}\right).
\end{equation}
Also, to obtain the temperature in this case we should first
evaluate the noncommutative surface gravity. Extending equation
(\ref{AA}) to the noncommutative spacetime (\ref{AC}), we get
\begin{equation}\label{AM}
\kappa_{nc}=\frac{\Lambda}{3}\hat{r},
\end{equation}
which should be evaluated at the noncommutative horizon radius
(\ref{AH}) with the result
\begin{equation}\label{AN}
\kappa_{nc}=\frac{\Lambda}{3}\sqrt{\frac{3}{\Lambda}-\theta^2}.
\end{equation}
Now, using the temperature formula (\ref{A}) in the noncommutative
case as well, we obtain the corresponding temperature as
\begin{equation}\label{AO}
T_{nc}=\frac{1}{2\pi}\frac{\Lambda}{3}\sqrt{\frac{3}{\Lambda}-\theta^2},
\end{equation}
or, up to second order of the noncommutative parameter
\begin{equation}\label{AP}
T_{nc}=T_{GH}\left(1-2\pi^2 \theta^2 T_{GH}^2\right).
\end{equation}
Finally, to calculate the vacuum energy of the noncommutative de
Sitter spacetime, we assume that the first law of thermodynamics
is also valid in this case as $dE_{nc}=T_{nc}dS_{nc}$. Use of
equations (\ref{AL}) and (\ref{AP}) then yields
\begin{equation}\label{AR}
dE_{nc}=dE_{GH}\left(1-\frac{1}{2}\frac{\theta^2}{E_{GH}^2}\right),
\end{equation}
or after integration
\begin{equation}
\label{AS}E_{nc}=E_{GH}\left(1+\frac{\theta^2}{2E_{GH}^2}\right).
\end{equation}
\section{Quantized entropy in noncommutative de Sitter
spacetime}In the previous section, to introduce noncommutativity
in de Sitter spacetime we directly replaced the coordinates with
their noncommutative counterparts. Subsequently, using
transformation (\ref{Z}) we found the radius (\ref{AI}) for the
noncommutative de Sitter space. In this section we shall deal with
this issue in a different manner by considering the relation
$r^2=r_h^2=3/\Lambda$ as a constraint. Assuming that this
constraint also holds in the noncommutative space in the form
$\hat{r}\hat{r}=r_h^2$, we proceed to quantize it through the use
of transformation (\ref{Z}) and use of the replacement
$p_i\rightarrow -i\partial_i$, that is
\begin{equation}\label{DA}
\left(x_i+\frac{i}{2}\Theta_{ij}\partial_j\right)
\left(x_i+\frac{i}{2}\Theta_{ik}\partial_k\right)\psi(\vec{x})=r_h^2\psi(\vec{x}),
\end{equation}
where $\psi(\vec{x})$ is a wave function which can be annihilated
by the constraint $\hat{r}\hat{r}-r_h^2=0$. As before, taking
$\Theta_{ij}=\epsilon_{ijk}\Theta_k$ and
$\vec{\Theta}=(0,0,\Theta)$, we obtain the following eigenvalue
equation
\begin{equation}\label{DB}
\left[-\frac{1}{2}\left(\frac{\partial^2}{\partial
x^2}+\frac{\partial^2}{\partial
y^2}\right)+\frac{2i}{\Theta}\left(x\frac{\partial}{\partial
y}-y\frac{\partial}{\partial
x}\right)+\frac{2}{\Theta^2}\left(x^2+y^2+z^2\right)\right]\psi(\vec{x})=
\frac{2r_h^2}{\Theta^2}\psi(\vec{x}),
\end{equation}
which is a Schr\"{o}dinger like equation for a charged particle
with unit mass moving in a magnetic field. Indeed, for a uniform
magnetic field $\vec{B}=(0,0,B)$, the Schr\"{o}dinger equation for
a particle with unit mass and charge $q$ reads
\begin{equation}\label{DC}
\left[-\frac{1}{2}\left(\frac{\partial^2}{\partial
x^2}+\frac{\partial^2}{\partial y^2}+\frac{\partial^2}{\partial
z^2}\right)+\frac{qB}{2}i\left(x\frac{\partial}{\partial
y}-y\frac{\partial}{\partial
x}\right)+\frac{q^2B^2}{8}\left(x^2+y^2\right)\right]\psi(\vec{x})=
E\psi(\vec{x}).
\end{equation}
As is well known from ordinary quantum mechanics the motion of
such a particle along the $z$-axis is a uniform motion
$z=v_{0z}t+z_0$ and the corresponding energy eigenvalues are the
Landau levels $E_n=\frac{1}{2}v_{0z}^2+qB(n+\frac{1}{2})$. Now, if
we take the constant $v_{0z}$ to be zero, the particle will move
in the plane $z=z_0$. In this case the wave function representing
the particle is independent of $z$ and the term
$-\frac{1}{2}\frac{\partial^2\psi}{\partial z^2}$ will be removed
from equation (\ref{DC}). Therefore, comparison of equations
(\ref{DC}) and (\ref{DB}) show that we can identify $qB$ with
$\frac{4}{\Theta}$ and bearing in the mind the Landau levels, we
find
\begin{equation}\label{DE}
\frac{2}{\Theta^2}(r_h^2-z_0^2)=\left(n+\frac{1}{2}\right)\frac{4}{\Theta}.
\end{equation}
To achieve the correct result in the limit $\Theta \rightarrow 0$
we take $z_0=\sqrt{3/\Lambda}$ and thus
\begin{equation}\label{DF}
r_h^2=\frac{3}{\Lambda}+2\Theta\left(n+\frac{1}{2}\right).
\end{equation}
Also, the entropy of the noncommutative de Sitter spacetime,
$S=A/4=4\pi r_h^2/4$, can be written as
\begin{equation}\label{DG}
S=\frac{3\pi}{\Lambda}+2\pi \Theta
\left(n+\frac{1}{2}\right).
\end{equation}
Therefore, we see that introduction of noncommutativity in this
approach yields certain quantization rules for the horizon radius
and entropy of the de Sitter spacetime.
\section{Thermodynamics of de Sitter spacetime with GUP}
Let us now study the model described above in the framework of GUP.
Here, the Heisenberg algebra is modified by the addition of small
corrections to the commutation relations. The resulting
modifications implies a minimum position uncertainty of order of the
Planck length suggested by quantum gravity and string theory. This
framework is thus suitable to the study of the de Sitter spacetime
at the quantum level.

We start with the GUP defined as
\begin{equation}\label{AT}
\delta x \delta
p\geq \hbar+l_p^2\frac{(\delta p)^2}{\hbar},
\end{equation}
where $l_p=\sqrt{G \hbar/c^3}$ is the Planck length. Thus,
uncertainty formula leads to \footnote{Although in units adopted
here the Planck length is equal to unity, we retain it in the
ensuing formulae to follow the role it plays which is an important
one.}
\begin{equation}\label{AU}
\delta p=\frac{\delta
x}{2l_p^2}\left[1-\sqrt{1-\frac{4l_p^2}{(\delta
x)^2}}\right].
\end{equation}
Now, applying the uncertainty principle to the horizon radius
results in
\begin{equation}\label{AV}
T\sim \frac{\delta p}{2 \pi}.
\end{equation}
Note that the same result is also valid for temperature of a black
hole, see \cite{1}. Taking $\delta x=r_h=\sqrt{3/\Lambda}$ and use
of equation (\ref{AU}) leads to the modified temperature based on
GUP
\begin{equation}\label{AX}
T_{GUP}=\frac{1}{4\pi
l_p^2}\sqrt{\frac{3}{\Lambda}}\left[1-\sqrt{1-\frac{4\Lambda
l_p^2}{3}}\right],
\end{equation}
where up to second order in Planck length we have
\begin{equation}\label{AY}
T_{GUP}=T_{GH}\left(1+4\pi^2 l_p^2
T_{GH}^2 \right).
\end{equation}
To evaluate the entropy of the de Sitter space in the GUP framework
we do as follows. Assume that because of particle radiation
described in \cite{17}, the area $A=4\pi r_h^2$ of the horizon
changes as
\begin{equation}\label{AZ}
dA=8\pi r_h dr_h,
\end{equation}
where $r_h=\sqrt{3/\Lambda}$. During this process the energy of
the de Sitter spacetime changes as $\delta E$ such that
\begin{equation}\label{BA}
\delta A=\frac{24\pi}{\sqrt{3\Lambda}}\delta E,
\end{equation}
where we have used relation (\ref{K}). Now, the uncertainty
principle for radiated particles implies $\delta E \sim \delta p
\sim 1/\delta x$ and thus
\begin{equation}\label{BC}
\delta A=\frac{24\pi}{\sqrt{3\Lambda}}\delta
p=\frac{24\pi}{\sqrt{3\Lambda}}\frac{1}{\delta x}.
\end{equation}
Extending this relation to the GUP case, we are
led to
\begin{equation}\label{BD}
\delta A_{GUP}=\frac{24\pi}{\sqrt{3\Lambda}}\frac{\delta
x}{2l_p^2}\left[1-\sqrt{1-\frac{4l_p^2}{(\delta x)^2}}\right],
\end{equation}
where keeping terms up to fourth order in Planck length yields
\begin{equation}\label{BE}
\delta A_{GUP}=\frac{24\pi}{\sqrt{3\Lambda}}\frac{1}{\delta
x}\left[1+\frac{l_p^2}{(\delta x)^2}+\frac{2l_p^4}{(\delta
x)^4}\right].
\end{equation}
Now, taking $\delta x=r_h=\sqrt{A/4\pi}$, we get
\begin{equation}\label{BF}
dA_{GUP}=\left(1+\frac{4\pi l_p^2}{A}+\frac{32\pi^2
l_p^4}{A^2}\right)dA.
\end{equation}
Integration then yields
\begin{equation}\label{BG}
A_{GUP}=A+4\pi l_p^2 \ln A -32\pi^2 l_p^4
\frac{1}{A}.
\end{equation}
Now, by extending the entropy formula $S=A/4$ to the GUP case we
obtain an expression for the entropy of the de Sitter space in the
GUP framework
\begin{equation}\label{BH}
S_{GUP}=S_{GH}+\pi l_p^2 \ln S_{GH}
-2\pi^2 l_p^4 \frac{1}{S_{GH}}+...+C,
\end{equation}
where $C$ is a constant proportional to $l_p^2$. Finally, to
obtain the energy of the de Sitter spacetime based on GUP, we
assume that the first law of thermodynamics is also valid in this
case : $dE_{GUP}=T_{GUP}dS_{GUP}$. Now, using equations (\ref{AY})
and (\ref{BH}), we find an expression for energy after integration
as
\begin{equation}\label{BI}
E_{GUP}=E_{GH}\left(1-\frac{2l_p^2}{E_{GH}^2}-\frac{l_p^4}{E_{GH}^4}\right).
\end{equation}

\section{Conclusions}
In this paper we have studied the effects of noncommutativity on the
thermodynamical properties of de Sitter spacetimes. The physical
relevance of such attempts is that de Sitter spacetimes radiate a
thermal spectrum which behaves like the Hawking radiation for black
holes and should be described quantum mechanically and thus includes
quantum gravitational effects, indicating that the spacetime is
noncommutative. We have used this analogy with the black hole
radiation formulae for temperature and entropy to obtain the
temperature of de Sitter spacetimes as a radiating thermodynamical
system and, therefore, assigned an entropy to it. Assuming that this
system obeys the first law of thermodynamics, we have found the
vacuum energy of de Sitter spacetimes. We then introduced
noncommutativity in the spacetime metric by replacing the
coordinates with their noncommutative counterparts. This led to new
relationships for radius of the horizon, temperature, entropy and
energy of the system, all showing similar results akin to those of
the commutative case but incorporating some correction terms. From
the noncommutative thermodynamical variables, we see that the
horizon radius, temperature and entropy of the de Sitter space are
decreased, while its energy increased.

The noncommutative coordinates was subsequently dealt with in an
alternative way. In this approach one can obtain the horizon radius
as a constraint equation which also holds in the noncommutative
case. Here, we have seen that the noncommutativity effect resembles
the presence of a magnetic field. Quantizing this constraint led us
to a quantization rule for the entropy of de Sitter spacetime. We
have also investigated the problem by applying GUP on the de Sitter
horizon. In this framework we have obtained the thermodynamical
variables with some different correction terms. Particulary, the
correction term in the entropy formula is a logarithmic term which
is a known correction term in quantum black hole physics. This is in
contrast to the previous result for noncommutative case in which the
first order correction term is a constant shift in (\ref{AL}).

An important issue in our analysis is that the different
approaches dealing with noncommutativity in a phenomenological
framework, result in different correction terms to thermodynamical
quantities of the de Sitter spacetime. For example $E_{nc}$ in
equation (\ref{AS}) is larger than $E_{GH}$ while $E_{GUP}$ in
equation (\ref{BI}) diminishes with respect to $E_{GH}$. This
difference in the sign of the correction terms also appears in the
black hole thermodynamics when one considers the quantum
gravitational effects \cite{31,32}. As is indicated in \cite{32}
``the sign of a logarithmic term can be understood as either
related to the missing information (if it is positive) or the
increase of information (if it is negative), depending on whether
the horizon area is fixed or not". A complete answer to this still
open question is not possible until a full theory of quantum
gravity emerges, a theory which presently does not exist in a
complete form.

\end{document}